\documentclass[a4paper,11pt]{article}
\usepackage{pos}
\usepackage{siunitx}
\usepackage{csquotes}
\usepackage{comment}
\usepackage{multirow}
\usepackage{lineno}
\renewcommand{\logo}{\relax} % removes PoS logo for arXiv submission

\DeclareSIUnit \solarmass {M_{\odot}}
\DeclareSIUnit \parsec {pc}
\DeclareSIUnit \erg {erg}

\title{Prospects for detecting periodic or sharp fast-time features in the supernova neutrino lightcurve with IceCube}
\ShortTitle{Feature detection in the CCSN neutrino lightcurve with IceCube}

\author*[a]{Jakob Beise}
\author[a]{María Durán de las Heras}
\author[b]{Segev BenZvi}
\author[b]{Spencer Griswold}
\author[c]{Nora Valtonen-Mattila}
\author[d]{Evan O'Connor}
\author[ef]{David Barba-González}
\author[a]{Erin O'Sullivan}

\affiliation[a]{Dept. of Physics and Astronomy, Uppsala University, Box 516, SE-75120 Uppsala, Sweden}
\affiliation[b]{Dept. of Physics and Astronomy, University of Rochester, NY 14627 Rochester, USA}
\affiliation[c]{Fakultät für Physik \& Astronomie, Ruhr-Universität Bochum, D-44780 Bochum, Germany}
\affiliation[d]{The Oskar Klein Centre, Dept. of Astronomy, Stockholm University,\\ AlbaNova, SE-10691 Stockholm, Sweden}
\affiliation[e]{Dept. of Fundamental Physics and IUFFyM, Universidad de Salamanca,\\ Plaza de la Merced s/n, E-37008 Salamanca, Spain}
\affiliation[f]{Racah Inst. of Physics, The Hebrew University, Jerusalem 9190401, Israel}

\emailAdd{jakob.beise@physics.uu.se}
\emailAdd{mariadurandelasheras@gmail.com}
\abstract{Neutrinos produced in core-collapse supernova offer a direct probe into the hydrodynamics and energy transport mechanisms during the collapse and play a pivotal role in the shock revival and success of the supernova explosion. Fast-time features of the neutrino luminosity and energy spectrum encode information about phenomena such as turbulence, convection, shock revival and potential quark-hadron phase transitions. In this study, we explore the detection capabilities of large-volume neutrino telescopes with a focus on IceCube and the planned extension IceCube-Gen2. Furthermore, we consider the effect on the detection sensitivity from wavelength shifters through enhanced light collection.

A variety of models predict periodic fast-time features in supernova light curves; to quantify their detectability without relying on specific models, we investigate the detector response to a generic parameterisation of such features. We find that independent of feature frequency, IceCube-Gen2 instrumented with wavelength shifters has sensitivity to weaker modulations ($>25\%$ amplitude) as compared to only the strongest modulations ($>50\%$ amplitude) with IceCube. In addition, we examine the sensitivity of the neutrino lightcurve to sharp features from a quark-hadron phase transition. Phase transitions leading to a quark star remnant are detectable with IceCube at $5\sigma$ up to the edge of the Galaxy, and throughout the Small Magellanic Cloud with IceCube-Gen2 equipped with wavelength-shifters. In contrast, models collapsing into a black hole are observable only within the Galaxy, covering 41\% of the CCSNe population for IceCube and nearly all (91\%) for IceCube-Gen2 complemented by wavelength shifters. These results highlight the potential of IceCube-Gen2 for detecting Galactic sources more reliably and with greater reach. 
}

\FullConference{19th International Conference on Topics in Astroparticle and Underground Physics (TAUP2025)\\
24–30 Aug 2025\\
Xichang, China\\}

\makeatletter
\g@addto@macro\normalsize{%
  \setlength{\abovedisplayskip}{6pt}%
  \setlength{\belowdisplayskip}{6pt}%
  \setlength{\abovedisplayshortskip}{4pt}%
  \setlength{\belowdisplayshortskip}{4pt}%
}
\makeatother

\begin{document}
\maketitle

\section{Introduction}
\label{sec:intro}

\noindent Core-collapse supernovae (CCSNe) are among the most violent astrophysical phenomena in the Universe and are key to understanding its chemical composition. Yet, the internal dynamics driving a successful explosion are poorly understood. Neutrinos, due to their weakly interacting nature, provide a direct probe of the internal conditions during collapse and play a critical role in transferring energy to the infalling material during the accretion phase.

Multi-dimensional simulations have established that neutrino-aided mechanisms, such as convective overturn in the neutrino-heated layer, the standing accretion shock instability (SASI) and rapid progenitor rotations \cite{Janka:2016anrv}, can revive the stalled shock by prolonging the dwell time of matter in the gain region. These mechanisms imprint periodic, fast-time modulations in the neutrino luminosity and energy spectrum on $\mathcal{O}(\SI{100}{\milli \s})$ timescales with frequencies ranging from \SI{50}{\Hz} to \SI{400}{\Hz}, linking the core dynamics to observable signals. IceCube's sensitivity strongly depends on the feature amplitude and progenitor mass, spanning roughly \SIrange{1}{35}{\kilo \parsec} \cite{Beise:2023sasi, Walk:2018snft, Walk:2020sasi, Westernacher-Schneider:2019snro}.

An alternative shock revival mechanism arises from the extreme conditions of the core shortly after core collapse. When the central density exceeds several times the nuclear density and the temperature rises above $\mathcal{O}(\SI{10}{\MeV})$, hadronic matter can undergo a phase transition to deconfined quark matter \cite{Fischer:2011core}. The resulting softening of the equation of state (EOS) triggers a second collapse $\approx 500-1000~\mathrm{ms}$ after the initial collapse producing a sharp burst of $\bar{\nu}_e$ of $\mathcal{O}(\SI{10}{\mega \eV})$ lasting \SIrange{0.1}{10}{\milli \s} \cite{Huang:2025phtr,Largani:2024cons}.

MeV neutrinos from CCSNe interact in long-string Cherenkov detectors like IceCube \cite{IceCube:2016inst} primarily via the inverse beta-decay (IBD) on protons in water or ice, making them most sensitive to the $\bar{\nu}_e$ flux. The resulting positrons travel $\mathcal{O}(\SI{10}{\centi \m})$, emitting Cherenkov photons typically detected by a single sensor due to the sparse detector geometry. With noise rates per sensor of up to \SI{540}{\Hz}, individual CCSN neutrino interactions cannot be reconstructed. Instead, IceCube would observe a Galactic CCSN as a collective rise of the detector rate lasting $\mathcal{O}(\SI{10}{\s})$ \cite{IceCube:2011pros}.

IceCube instruments \SI{1}{\kilo \m^3} of clear, glacial ice with 5160 digital optical modules (DOMs) on 86 cables (\enquote{strings}) at depths of \SI{1450}{\m} to \SI{2450}{\m} below the surface. Each optical sensor consists of a single, downwards-facing photomultiplier tube (PMT) inside a pressure housing. The IceCube Upgrade \cite{Ishihara:2019}, currently being installed, adds about 700 new, densely spaced sensors surrounding the inner DeepCore region, improving detector calibration and enabling precise neutrino oscillation measurements. IceCube-Gen2 \cite{IceCube-Gen2:2023tdr} is a planned large-scale extension of IceCube, adding some 10,000 new optical modules on 120 new \enquote{strings}. These modules will be segmented, each containing multiple, small-diameter PMTs with nearly $4\pi$ coverage. Incorporating wavelength-shifting (WLS) modules can further enhance performance in low signal-to-noise regimes, such as CCSN detection, through augmented photo collection and reduced noise. A demonstrator, the Wavelength-shifting Optical Module (WOM) \cite{WOM:2022ini}, will be tested in the IceCube Upgrade. WLS tubes coupled to segmented sensors are a cost-efficient, low-noise and scalable designed under consideration in IceCube-Gen2.

%IceCube will detect CCSNe across the Milky Way (MW), model-independently at $>10\sigma$ and has sensitivity for heavier progenitors in the Large and Small Magellanic Cloud (LMC and SMC) \cite{IceCube:2011pros, IceCube:2024ccsn}. IceCube-Gen2's segmented sensors have an increased sensor photocathode area, which, in combination with coincidence triggers for noise reduction, would extend the sensitivity to low-mass progenitors in the LMC and SMC \cite{IceCube-Gen2:2023tdr, LozanoMariscal:2021mult}.
Due to the enormous statistics from a Galactic supernova, IceCube and its planned extensions can not only detect the neutrino lightcurve but also extract valuable physics from the details therein. This contribution presents the latest results for detecting generic, periodic fast-time features as well as sharp features from quark-hadron phase transitions.

\section{Analysis}
\label{sec:ana}

\noindent The number of detected signal photons in a time interval $\Delta t$ is obtained by summing over all sensor types $i$, each contributing with $m_i$ sensors, multiplied by the time-integrated, average per-sensor hit rate $\langle R_{\mathrm{SN}, i} (t)\rangle$ (see Ref.~\cite{IceCube:2011pros}):
\vspace{-4mm}
\begin{equation}
    \label{eq:sig}
     N_{\mathrm{sig}} = \sum_i m_i \int_{0}^{\Delta t} dt \ \langle R_{\mathrm{sig}, i} (t) \rangle \ .
\end{equation}
%\vspace{-0mm}
\noindent The average single-sensor hit rate for a specific reaction and target is defined as
%
%\vspace{-3mm}
\begin{equation}
    \langle R_{\mathrm{sig}, i}(t)\rangle =  \langle \epsilon_{\tau, i} \rangle \times \frac{n_{\mathrm{tar}} \  \mathcal{L}_{\nu}}{4\pi d^2 \langle E_{\nu} \rangle} \int_{0}^{\infty} dE_{e} \int_{0}^{\infty} dE_{\nu}
    \times \frac{d \sigma}{d E_{e}} \ N_{\gamma}(E_{e})
    \times V_{\gamma, i}^{\mathrm{eff}} \ f(E_{\nu}) \ ,
\end{equation}
\noindent where $\langle \epsilon_{\tau, i} \rangle$ is the dead time efficiency, $n_{\mathrm{tar}}$ is the density of targets in ice, $d$ is the distance to the supernova with neutrino luminosity $\mathcal{L}_{\nu}(t)$ and normalised neutrino energy distribution $f(E_{\nu})$. $V_{\gamma, i}^{\mathrm{eff}}$ is the effective volume per sensor. $N_{\gamma}(E_{e})\approx 178 \cdot E_{e}$ is the energy-dependent number of radiated Cherenkov photons and $\frac{d \sigma}{d E_{e}}$ is the differential cross section, while $E_\nu$ ($E_e$) is the neutrino (positron) energy. The time dependence arises through $\mathcal{L}_\nu(t)$ and $f(E_\nu, t)$. The number of background hits over a time $\Delta t$ for sensors with non-paralysing dead time $\tau$ is the sum over all sensors with noise rate $R_\tau$:
\vspace{-4mm}
\begin{equation}
    \label{eq:bkg}
    N_{\mathrm{bkg}} = \sum_i m_i \ \Delta t \  \langle R_{\tau, i} \rangle \ .
\end{equation}
We consider the detector geometries listed in Tab.~\ref{tab:detector} and use the analytic detector response calculator \texttt{ASTERIA} \cite{ASTERIA} to compute $N_\mathrm{sig}$ and $N_{\mathrm{bkg}}$. In both studies, we evaluate the discriminating power between a \enquote{flat} lightcurve (null hypothesis, $H_0$) and one that exhibits either periodic or sharp fast-time features (signal hypothesis, $H_1$).

\begin{table}
    \caption{Characteristics of detector geometries considered in this study. IceCube comprises 4800 standard DOMs and 360 high-quantum efficiency DOMs in the DeepCore region.\label{tab:detector}}
    \centering
    \begin{tabular}{lcccc}
    %\hline\hline
    Geometry & $m$ & $\langle V_{\gamma}^{\mathrm{eff}} \rangle$ [\SI{}{\m^{3}}] & $\langle R_{\tau} \rangle$ [Hz]\\
    \hline
    \multirow{2}{*}{IceCube} & 4800 &  0.17 & 285\\
    & 360 & 0.23 & 359\\
    \hline
    Gen2 & 9760 & 0.33 & 2300\\
    Gen2+WLS & 9760 & 0.60 & 2700\\
    %\hline
    \end{tabular}
\end{table}

\subsection{Generic, periodic features}
\label{sec:ana_generic}

\noindent Here, we summarise the main findings regarding the detection of generic, periodic fast-time features published in Ref.~\cite{Beise:2025gene}. We use the \SI{27}{\solarmass} model from Ref.~\cite{Sukhbold:2016ccsn} with LS220 EOS as a baseline to compute the number of hits under $H_0$, $\mathcal{X}_{H_0}$. The model was chosen because it does not exhibit fast-time features and predicts a moderate detection rate, compared to models of similar progenitor mass. We consider a total signal duration of \SI{1000}{\milli \s} and inject a full-integer, \SI{150}{\milli \s}-long sinusoidal modulations (defined between \SIrange{150}{300}{\milli\s}) with frequency $f \in [50,~400]~\mathrm{Hz}$ and amplitude $A'$ relative to the maximum of $\mathcal{X}_{H_0}$, $A' \in [2.5,10,2.5]\%\ \cup [10,50,5]\%$. We apply a Hann window to the fast-time feature to remove edge discontinuities. We considered the optimistic scenario of no neutrino flavour mixing at source or Earth, which we review together with other modelling uncertainties in Sec.~\ref{sec:res_generic}.

The maximum power of the time-dependent, short-time fourier transform (STFT) of $\mathcal{X}_{H_0}$ and $\mathcal{X}_{H_1}$, $\widetilde{\mathcal{X}}_{H_0}$ and  $\widetilde{\mathcal{X}}_{H_1}$, is used as a test statistic (TS) to distinguish $H_1$ from $H_0$. The STFT is applied on a \SI{100}{\milli \s}-long extract, sliding through the full signal in strides of \SI{20}{\milli \s} resulting in a time (frequency) resolution of \SI{20}{\milli \s} (\SI{10}{\Hz}). A $\SI{50}{\Hz}$ high-pass filter is applied to reject low-frequency artefacts from the overall shape of lightcurve. Due to the non-Gaussian tails of the background TS, $\mathrm{TS}_{H_0}$, we pre-generate high-statistics ($n_{\mathrm{bkg}}=10^8$) trials on a \SI{0.2}{\kilo \parsec} distance grid. Since the modes of $\mathrm{TS}_{H_1}$ are statistically stable, we only run $n_{\mathrm{sig}}=10^4$ signal trials. We directly compute the $p$-value of the separation from data and use a one-sided Gaussian test to obtain the detection significance $\xi$. The detection horizon is the distance $d_\xi$ for which $\xi$ falls below a detection threshold, e.g. $3\sigma$ (evidence) or $5\sigma$ (discovery) which we compute for all 420 parameter combinations ($f$, $A'$).

\subsection{Sharp features}
\label{sec:ana_sharp}

\noindent In addition to the generic case above, we consider phase transition-driven core collapse models. We simulate in GR1D \cite{OConnor:2015grid} and Sedonu \cite{Gullin:2022sedo} the evolution of two models from \cite{Largani:2024cons}.  First, the s30a28 RDF-1.2 model, resulting in a successful explosion forming a quark star remnant, and the s40a28 RDF-1.7 model, which has a delayed collapse into a black hole after the phase transition. Both models differ in the $\bar{\nu}_e$ peak luminosity and duration of the second collapse some $\approx\SI{1}{\s}$ after core bounce, with $\mathcal{L}_{\bar{\nu}_e}^{\mathrm{max}} \approx \SI{2.5e53}{\erg/\s}$ (\SI{1e52}{\erg/\s}) over \SI{11}{\milli \s} (\SI{6}{\milli \s}) for the RDF-1.2 (RDF-1.7) model. We make the optimistic assumption of no flavour mixing at source.

For a detailed description of the analysis method, we refer to Ref.~\cite{Duran:2025pros}. In contrast to the methodology introduced in Sec.~\ref{sec:ana_generic}, the feature duration is extremely short, and the narrow peak would be split over multiple frequency bins, diminishing our sensitivity. We instead opt for a time domain-based, maximum amplitude analysis. Consequently, we adapt a much finer temporal resolution of \SI{0.05}{\milli \s} (\SI{0.1}{\milli \s}) for the RDF-1.2 (RDF-1.7) model to match the full-width half-maximum of the $\bar{\nu}_e$ burst. This is feasible in practice using IceCube's nano-second resolution data stream \texttt{HitSpool} data stream \cite{Zuydtwyck:2015hisp}. We obtain the \enquote{smooth} lightcurve under $H_0$, by applying a low-pass filter with a \SI{50}{\Hz} (\SI{500}{\Hz}) cut-off frequency for the RDF-1.2 (RDF-1.7) model. The significance of any observed deviation from $H_0$ is evaluated using a TS test, defined by the maximum amplitude in the \SI{11}{\milli \s} (\SI{6}{\milli \s}) signal window.
Following the same method as for the generic, periodic features, we pre-generate high-statistics background trials ($n_{\mathrm{bkg}}=10^8$) and evaluate the detection significance with $n_{\mathrm{sig}}=10^4$ signal trials.

\section{Results}
\label{sec:res}

\subsection{Generic, periodic features}
\label{sec:res_generic}

\begin{figure}
    \centering
    \includegraphics[width=0.8\linewidth]{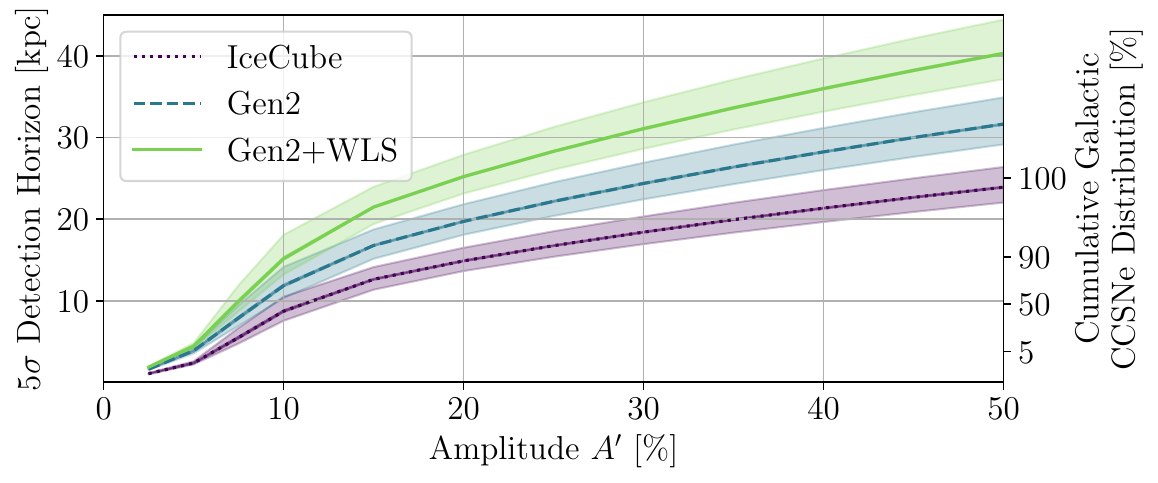}
    \caption{Frequency averaged $5\sigma$ detection horizon as a function of the relative amplitude $A'$ of the generic fast-time feature for the detector geometries from Tab.~\ref{tab:detector}. The error bands indicate the 68\% containment interval. The \textit{right y-axis} indicates the Galactic CCSN coverage following the parametrisation from Ref.~\cite{Adams:2013rate}.}
    \label{fig:result_generic}
\end{figure}

\noindent We find no dependence for frequency on the detection capability with the injected frequency $f$; hence, we define the frequency-averaged detection horizon $\langle d_\xi \rangle$, plotted as a function of the relative amplitude $A'$ in Fig.~\ref{fig:result_generic} for the $5\sigma$ detection threshold. The \textit{right y-axis} illustrates the Galactic CCSNe coverage from \cite{Adams:2013rate} assuming all Galactic CCSN have the same properties as the Sukhbold 2015 model. We find that Gen2+WLS is sensitive at the $5\sigma$ (3$\sigma$) level to periodic fast-time features with amplitudes $A' \geq 25\%$ $(\geq 20\%)$ throughout the Milky Way, as compared to IceCube's limited sensitivity to the strongest modulations probed ($A' \geq 50\%$).

We find that more realistic neutrino flavour conversion scenarios, such as resonant Mikheev-Smirnov-Wolfenstein (MSW) flavour conversion at source reduce the detection horizon but at most 16\%. Furthermore, a shorter, \SI{50}{\milli \s}-long, periodic feature is visible to half the distance as our reference \SI{150}{\milli \s}-long signal. Drifts in the feature frequency diminish the sensitivity, with a \SI{2}{\Hz / \milli \s} reducing the detection horizon by a factor of 2. In addition, two-frequency modes (beats) barely affect our results, as long as the beat frequency is larger than the time resolution.

\subsection{Sharp features}
\label{sec:res_sharp}

\begin{figure}
    \centering
    \includegraphics[width=0.8\linewidth]{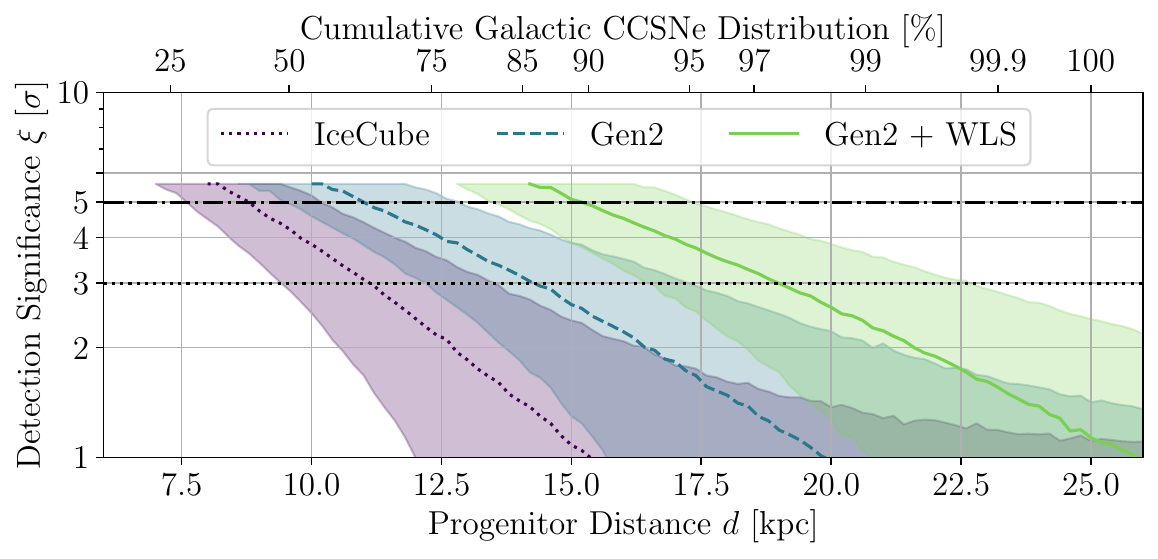}
    \caption{Detection significance $\xi$ as a function of progenitor distance $d$ for the RDF-1.7 quark-hadron phase transition model for the detector configurations from Tab.~\ref{tab:detector}. The error bands indicate the 68\% containment interval. We also indicate the Galactic CCSN coverage from Ref.~\cite{Adams:2013rate} on the \textit{top axis}.}
    \label{fig:result_qcd}
\end{figure}

\noindent In Fig.~\ref{fig:result_qcd}, we present the detection significance for the model collapsing into a black hole (RDF-1.7). The sharp feature can be detected at $5\sigma$ ($3\sigma$) level to distances of $8.8^{+1.5}_{-1.2}~\mathrm{kpc}$ ($11.1^{+2.4}_{-1.7}~\mathrm{kpc}$) in IceCube, $11.0^{+1.8}_{-1.5}~\mathrm{kpc}$ ($14.3^{+3.1}_{-2.1}~\mathrm{kpc}$) in Gen2 and $15.2^{+2.2}_{-1.7}~\mathrm{kpc}$ ($19.0 \ ^{+3.7}_{-2.6}~\mathrm{kpc}$) in Gen2+WLS, the latter covering up to 90\% of the Milky Way up from 41\% with IceCube alone. IceCube has higher sensitivity to the quark star remnant model (RDF-1.2), owing to the factor $\approx 25$ larger $\bar{\nu}_e$ flux. The sharp feature is detectable at $5\sigma$ throughout the Galaxy and up to the LMC and SMC for Gen2 and Gen2+WLS, respectively.

\begin{comment}
RDF 1.7 (BH)
IceCube reaches 3.0σ at 11.11 kpc (+2.43/−1.70)
IceCube reaches 5.0σ at 8.80 kpc (+1.46/−1.20)
Gen2 reaches 3.0σ at 14.25 kpc (+3.05/−2.09)
Gen2 reaches 5.0σ at 11.00 kpc (+1.75/−1.50)
Gen2+WLS reaches 3.0σ at 18.98 kpc (+3.67/−2.57)
Gen2+WLS reaches 5.0σ at 15.15 kpc (+2.18/−1.68)

RDF 1.2 (QS)
    IceCube reaches 3.0σ at 46.77 kpc (+7.82/−5.48)
IceCube reaches 5.0σ at 39.22 kpc (+4.72/−3.65)
Gen2 reaches 3.0σ at 60.84 kpc (+9.71/−7.16)
Gen2 reaches 5.0σ at 50.83 kpc (+5.71/−4.64)
Gen2+WLS reaches 3.0σ at 78.36 kpc (+12.35/−8.73)
Gen2+WLS reaches 5.0σ at 65.79 kpc (+7.14/−5.60)
\end{comment}

\section{Conclusion}
\label{sec:concl}
\noindent We presented the first, model-independent study of periodic fast-time features in neutrino detectors. Compared to IceCube, Gen2+WLS has $5\sigma$ sensitivity throughout the Milky Way to models with amplitudes larger than 25\%, down from extremely optimistic models of 50\% amplitude variation measurable in IceCube. 
For quark-hadron phase transition models that successfully explode, Gen2+WLS extends the detection capability beyond the edge of the Milky Way, well containing CCSNe explosions in the LMC and SMC. The abrupt neutrino cessation from models collapsing into a black hole limits the sensitivity to the Milky Way, with Gen2+WLS increasing Galactic coverage by a factor of 2 with respect to the current IceCube setup. 
Our findings underline that IceCube-Gen2 (equipped with WLS technology) will constitute a precision instrument to study even faint features in the lightcurve of Galactic CCSNe.

\section{Acknowledgements}

\noindent\small{The authors gratefully acknowledge grant no. 2019-05447 by the Swedish Research Council. The computations and data handling were enabled by resources provided by the National Academic Infrastructure for Supercomputing in Sweden (NAISS), partially funded by the Swedish Research Council through grant agreement no. 2022-06725.}

%\vspace{-3mm}
\bibliographystyle{JHEP}
\small{
\bibliography{references}
}
%\begin{thebibliography}{99}
%\bibitem{...}
%....
%
%\end{thebibliography}

\end{document}